\documentclass{aa}
\usepackage{graphicx}
\usepackage{txfonts}
\usepackage{xcolor}
\usepackage[switch]{lineno}
\begin{document}

%-----------------------------------------------------------------------
% Title, Authors, and Abstract
\title{ConKer: evaluating isotropic correlations of arbitrary order}

\author{Z. Brown
\and G. Mishtaku
\and R. Demina}

\institute{Department of Physics and Astronomy, University of Rochester,\\
500 Joseph C. Wilson Boulevard, Rochester, NY 14627, USA\\
\email{zbrown5@ur.rochester.edu}}

\date{Received XX XX, XXXX; accepted YY YY, YYYY}

\abstract{High order correlations in the cosmic matter density have become increasingly valuable in cosmological analyses. However, computing such correlation functions is computationally expensive.}{We aim to circumvent these challenges by designing a new method of estimating correlation functions.}{This is realized in {\tt ConKer}, an algorithm that performs $FFT$ convolutions of matter distributions with spherical kernels.}{{\tt ConKer} is applied to the CMASS sample of the SDSS DR12 galaxy survey and used to compute the isotropic correlation up to correlation order $n=5$. We also compare the $n=2$ and $n=3$ cases to traditional algorithms to verify the accuracy of the new method. We perform a timing study of the algorithm and find that two of the three components of the algorithm are independent of the catalog size, $N$, while one component is $\mathcal{O}(N)$, which starts dominating for catalogs larger than 10M objects. For  $n<5$  the dominant calculation is $\mathcal{O}(N_c^{4/3} \log N_c)$, where $N_c$ is the number of the grid cells.  For higher $n$, the execution time is expected to be dominated by a component with time complexity $\mathcal{O}(N_c^{(n+2)/3})$.}{We find {\tt ConKer} to be a fast and accurate method of probing high order correlations in the cosmic matter density. }


\maketitle
%-----------------------------------------------------------------------
% Introduction
\section{Introduction}
\label{sec:intro}

Understanding the dynamics of inflation in the early universe is tied with the study of primordial density fluctuations, in particular with their deviations from a Gaussian distribution (see e.g. \cite{Maldacena:2002vr}, \cite{Bartolo:2004if}, \cite{Acquaviva:2002ud}).
High order correlations have been shown to be sensitive to non-Gaussian density fluctuations  (see \cite{Meerburg:2019qqi} and references within). Yet, a brute force approach leads to prohibitively expensive $\mathcal{O}(N^n)$ computation of correlations, where $N$ is the number of objects and $n$ is the correlation order. This problem has been studied extensively~\citep{MarchThesis}. Several approaches to mitigate it were suggested for the calculation of three point correlations, each relying on a particular set of assumptions, e.g. small angle \citep{Yuan:2018qek}, Legendre expansion \citep{Slepian:2015qza}. The later approach was recently generalized for the $n$-point correlation function in~\cite{philcox2021encore} resulting in an $\mathcal{O}(N^2)$ algorithm. Here, we present an alternative computationally efficient way of evaluating such correlations. Similar to \cite{MarchThesis}, this algorithm exploits spatial proximity and similar to \cite{zhang2011FFT} and \cite{Slepian:2015qwa}, it uses a Fast Fourier Transformation ($FFT$). These characteristics combined with implementation facilities help achieve a notable reduction in calculation time without much memory overhead. The developed algorithm is applicable to discrete matter tracers, such as galaxies, as well as to continuous ones, such as Lyman-$\alpha$ and 21 cm line intensity, or matter maps derived from weak lenses. The method can be applied to evaluate self-correlations, as well as cross-correlations between different matter tracers. The algorithm {\it Convolves Kernels} with matter maps, hence it is named {\tt ConKer}\footnote{The {\tt python3} implementation of this algorithm may be downloaded at \url{https://github.com/mishtak00/conker}}. It is an extension of the {\tt CenterFinder} algorithm \citep{brown2021algorithm}, designed to find locations in space likely to be the centers of the Baryon Acoustic Oscillations.

%-----------------------------------------------------------------------
% Algorithm description
\section{Algorithm description} 
\label{sec:description}

%-----------------------------------------------------------------------
% Input
\subsection{Strategy} 
\label{subsec:Strategy}

\begin{figure}
\includegraphics[width=\linewidth]{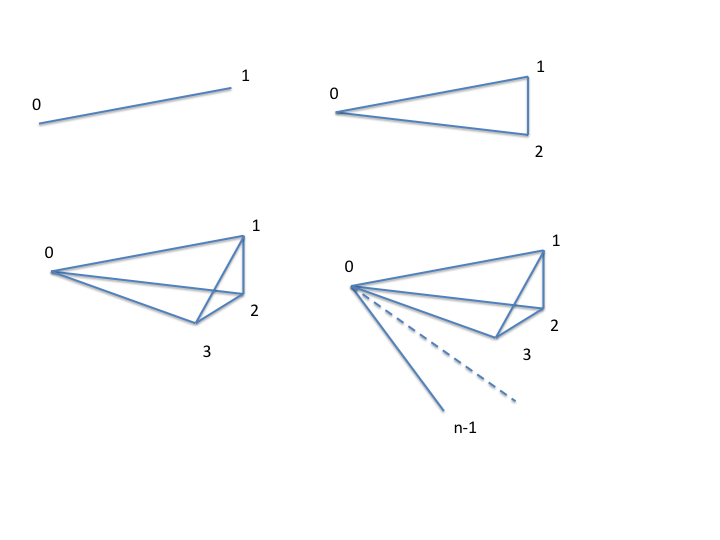}
\vspace{-1.7cm}
\caption{A cartoon illustrating two point correlation ($0-1$), three point correlation ($0-1-2$), four point correlation ($0-1-2-3$), and $n$  point correlation ($0-1-2-3-...-(n-1)$).}
\label{fig:n-pletes}
\end{figure}

Let $\rho(X,Y,Z)$ be the density of the matter tracer (e.g. galaxy count per unit volume) at a location with Cartesian coordinates $(X,Y,Z)$, with the $\bar{\rho}(X,Y,Z)$ being the density of expected observations from tracers randomly distributed over with surveyed volume. We define the deviation from the average density as

\begin{align}
\Delta(X,Y,Z) = \rho(X,Y,Z)- \bar{\rho}(X,Y,Z)\ . 
\label{eq:delta_def}
\end{align}

The two point correlation can be visualized as an excess of sticks of a given length over the ones that are randomly distributed over space. The three point correlation corresponds to an excess of triangles, the four point correlation - to counting pyramids (the four points do not necessarily lie in one plane), etc (see Fig~\ref{fig:n-pletes}). We will refer to these figures as $n$-pletes. 

Let us consider all possible $n$-pletes with one vertex at point $0$, characterized by a vector $\vec{r}$, with sides radiating from this vertex equal to $s_1,s_2, ... s_{(n-1)}$. In the isotropic case considered here the lengths of other sides and all angles are arbitrary. 

\begin{figure}
\includegraphics[width=\linewidth]{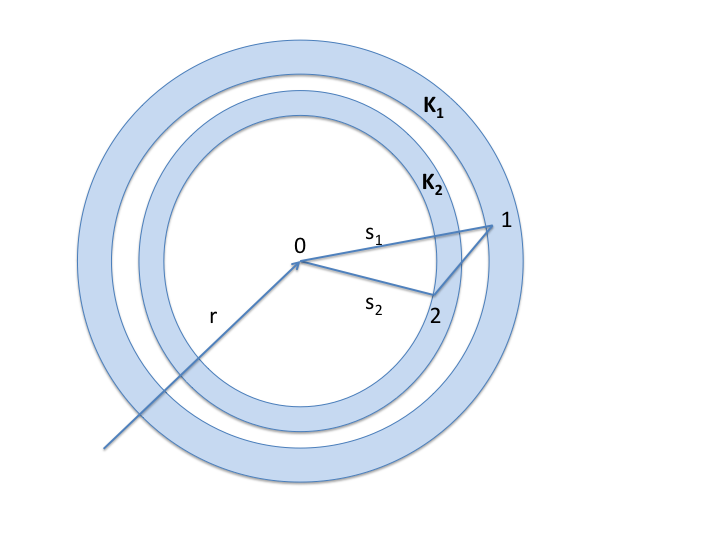}
\vspace{-0.8cm}
\caption{A cartoon illustrating  three point correlation ($0-1-2$) with different scales $s_1$ on the side ($0-1$) and $s_2$ on the side ($0-2$). Integration over spherical shells $K_1$ and $K_2$ is equivalent to counting all possible triangles that have one vertex at point $0$ and the other two anywhere on the spheres.}
\label{fig:ConKer_K1_K2}
\end{figure}

The isotropic $n$-point correlation function ($n$pcf) is defined as

\begin{multline}
\xi_n(s_1,s_2,...s_{(n-1)}) = \\ 
\frac{\int \Delta(\vec{r}) \Delta(\vec{r}+\vec{s_1}) \Delta(\vec{r}+\vec{s_2}) ...\Delta(\vec{r}+\vec{s}_{(n-1)}) d\vec{r} d\Omega_1 d\Omega_2... d\Omega_{(n-1)}} {\int \bar{\rho}(\vec{r}) \bar{\rho}(\vec{r}+\vec{s_1}) \bar{\rho}(\vec{r}+\vec{s_2}) ...\bar{\rho}(\vec{r}+\vec{s}_{(n-1)}) d\vec{r} d\Omega_1 d\Omega_2... d\Omega_{(n-1)}} \ . 
\label{eq:corr_theory}
\end{multline}
where the integration over $\vec{r}$ implies all possible positions of point $0$ in the surveyed volume.  The integration over a solid angle $d\Omega_i$($i=1,...n-1$) between vectors $\vec{s_i}$ and $\vec{r}$ is equivalent to integration over a sphere of  radius $s_i$ centered on point $0$, which, thus, evaluates the total number of tracers displaced from point $0$ by the distance $s_i$. This is illustrated in Fig.~\ref{fig:ConKer_K1_K2} for three point correlation. In term of Legendre polynomials expansion this definition of $n$pcf corresponds to $l=0$.

{\tt ConKer} evaluates the integral of the $\Delta$ field on a spherical shell of radius $s_i$  by placing a spherical kernel $K_i$ on point $0$ with Cartesian coordinates $(X,Y,Z)$ and taking its inner product with the $\Delta$ field. Then the kernel is moved to a different location, thus scanning the entire surveyed volume. The results of each step, $W_i(X,Y,Z)$, is a map of weights that characterize the integral of the density deviation on a spherical shell of radius $s_i$ around the location $(X,Y,Z)$. 

%-----------------------------------------------------------------------
% Input
\subsection{Input} 
\label{subsec:input1}

The inputs to the algorithm are catalogs of the observed number count of tracers $D$, with the total number being $N_{tot}$, and $R$, which represents a number count of randomly distributed points within the same fiducial volume. Most surveys provide the angular coordinates: right ascension $\alpha$ and declination $\delta$, and the redshift $z$ of the tracer. The relationship between the redshift and the comoving radial distance is cosmology dependent:

\begin{align}
r(z) = \frac{c}{H_0} { \displaystyle\int_{0}^{z} } \frac{dz'}{ \sqrt{\Omega_M(z'+1)^3+\Omega_k(z'+1)^2+\Omega_\Lambda} }\ , 
\label{eq:hubbleint}
\end{align} 
where $\Omega_M$, $\Omega_k$, and $\Omega_\Lambda$ are the relative present day matter, curvature, and cosmological constant densities respectively. $H_0$ is the present day Hubble's constant, $c$ is the speed of light. These user-defined parameters represent the fiducial cosmology. The integral in Eq.~\ref{eq:hubbleint} is evaluated numerically in {\tt ConKer}.
The comoving Cartesian coordinates are computed with:

\begin{align}
X=r \cos(\delta)\cos(\alpha)\ , 
\label{eq:cartesianx} \\
Y=r \cos(\delta)\sin(\alpha)\ ,
\label{eq:cartesiany} \\
Z=r \sin(\delta)\ .
\label{eq:cartesianz}
\end{align}
Cartesian coordinates are given as capital letters, $X$, $Y$, $Z$, as the redshift, $z$, is denoted in lower case.

We define a grid with spacing $g_s$, such that the volume of each cubic grid cell is $g_s^3$. $(X,Y,Z)$ denotes the Cartesian coordinates of the center of a grid cell. On this grid we define 3-dimensional histograms $D(X,Y,Z)$ and $R(X,Y,Z)$, which represent tracer counts in the cell $(X,Y,Z)$ from data and random catalogs respectively. These histograms may be populated by the raw count, or weighted count of tracers from the input catalogs. To properly account for the variation in angular completeness and the redshift selection function, the expected count from the random distribution, normalized to $N_{tot}$, is subtracted from the galaxy survey $D$ on the cell-by-cell basis. $R$ is evaluated using a random catalog. As an alternative, $R$ could be evaluated assuming that the tracers' angular, $P_{ang}(\alpha,\delta)$ and redshift\footnote{The redshift selection function $P_z(z)$ does not need to be uniform over $(\alpha,\delta)$, it could be defined piece-wise.}, $P_z(z)$ probability density distributions are factorizable \citep{demina2018computationally}. Then the expected count from the random distribution of galaxies in a Cartesian cell is:

\begin{align}
R(X,Y,Z) = N_{tot} [ P_{ang}(\alpha,\delta) \times P_z(z) ] \frac{dXdYdZ}{dV_{sph}} \ , 
\label{eq:factorizablility}
\end{align} 
where $dXdYdZ$ and $dV_{sph}=r^2d \sin \delta d \alpha$ give the volume of a particular cell in Cartesian and spherical sky coordinates.
This method of estimating the expected count from the random distribution reduces the statistical fluctuations compared to the count taken directly from the random catalog. Both methods are realized in the algorithm and may be chosen by the user. 
 
A 3-dimensional local density variation histogram $N(X,Y,Z)$, which is a discretized representation of the $\Delta$ field, is then defined on the grid to represent the difference in counts between $D$ and $R$:

\begin{align}
N(X,Y,Z) = D(X,Y,Z) - R(X,Y,Z)\ . 
\label{eq:bgsubtracted_map}
\end{align}
This stage of the algorithm is referred to as \textit {mapping}. 

%-----------------------------------------------------------------------
% Kernels
\subsection{Kernels} 
\label{subsec:kernel}

We construct $(n-1)$ spherical kernels, $K_i(X,Y,Z)$ on a cube of size $2s_i$ in all directions, just large enough to encompass a sphere of radius $s_i$. The grid defined on this cube has the same spacing, $g_s$, as the one used to construct $D(X,Y,Z)$. Grid cells intersected by a sphere of radius $s_i$  centered on the center of the cube, are assigned a value of 1. All other cells are 0.

%-----------------------------------------------------------------------
% The Convolution
\subsection{Convolution} 
\label{subsec:conv}

\begin{figure}
\includegraphics[width=\linewidth]{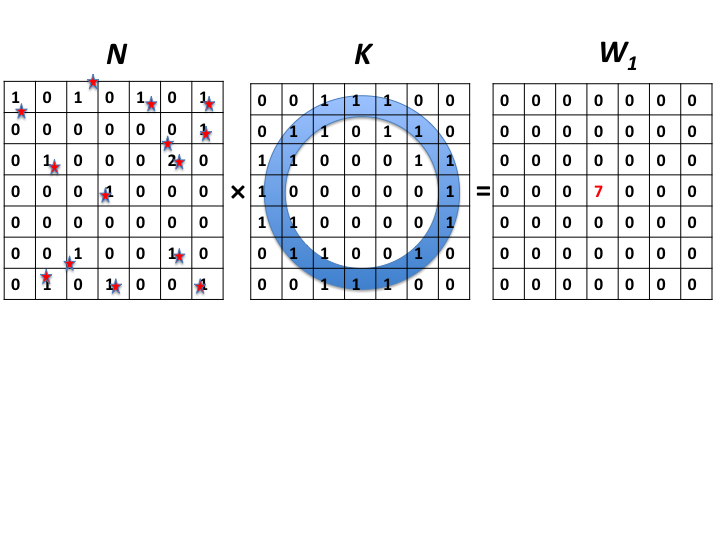}
\vspace{-3.1cm}
\caption{A 2-dimensional representation of one step in the convolution between  the matter histogram $N$ (left) and the spherical kernel $K$ (center).  In this example, the inner product is 7, which is placed in the center cell of $W_1$ histogram (right).}
\label{fig:cf_method}
\end{figure}

We construct a 3-dimensional histogram $W_i(X,Y,Z)$ with entries  equal to the inner product of kernel $K_i$  centered on a cell  with coordinates $(X,Y,Z)$ and matter density variation field $N$: 

\begin{align}
W_i(X,Y,Z) \equiv N(X,Y,Z)\ast\ast\ast K_i(X,Y,Z) \ , 
\label{eq:countconv}
\end{align} 
where $\ast\ast\ast$ denotes a 3-dimensional discrete convolution. 
One step in this process is visualized in Fig.~\ref{fig:cf_method}. 

This procedure is realized using an $FFT$ in a highly vectorized algorithm. In this process, all $N(X,Y,Z)$ cells are interpreted once as center points, i.e. point 0 in Fig. 2, and other times as points on shells. Thus the 3-dimensional histogram $W_0(X,Y,Z)$ is approximated as simply being the density variation in each cell of $N(X,Y,Z)$, i.e. $W_0=N$. 

For normalization purposes, we perform the same procedure on the field of random counts. The result of the convolution of random density field $R$ with kernel $K_i$ is referred to as $B_i(X,Y,Z)$:

\begin{align}
B_i(X,Y,Z) \equiv R(X,Y,Z)\ast\ast\ast K_i(X,Y,Z) \ , 
\label{eq:randconv}
\end{align} 
This stage of the algorithm is referred to as \textit {convolution}.

%-----------------------------------------------------------------------
% n-point correlation function
\subsection{n-point correlation function}

The $n$pcf, which approximates the one defined in Eq.~\ref{eq:corr_theory} is calculated as:

\begin{align}
\xi_n(s_1,s_2...s_{(n-1)}) =\frac{\sum W_0  W_1 W_2...W_{(n-1)}}{\sum B_0  B_1 B_2...B_{(n-1)} } \ . 
\label{eq:ksin_delta}
\end{align} 
The 2pcf is then

\begin{align}
\xi_2(s_1) = \frac{\sum W_0  W_1}{\sum B_0  B_1}  \ . 
\label{eq:ksi2_delta}
\end{align} 
The 3pcf necessitates convolution with kernel $K_2$ of radius $s_2$, resulting in weights $W_2$. The 3pcf is then calculated as 

\begin{align}
\xi_3(s_1,s_2) =\frac{\sum W_0  W_1 W_2} {\sum B_0  B_1 B_2}\ . 
\label{eq:ksi3}
\end{align}
The stage of the evaluation of the $n$pcf is referred to as \textit {summation}.

%-----------------------------------------------------------------------
% Equidistant case
\subsection{Equidistant case}

Of particular interest are correlations between objects with a defined scale, such as those  arising from a spherical sound wave in the primordial plasma, a.k.a. baryon acoustic oscillation. In this case, one tracer taken as a starting point is displaced from the rest $(n-1)$ points  by the same distance $s_1$, e.g. the three point correlation  corresponds to isosceles triangles randomly distributed in space. This equidistant case corresponds to the diagonal of the $n$-point correlation function, which is calculated as 

\begin{align}
\xi_n^{diag}(s_1) =\frac{\sum W_0  W_1^{n-1}} {\sum B_0  B_1^{n-1}}\ . 
\label{eq:ksinDiag}
\end{align} 

Note that only $W_0$, $W_1$, $B_0$ and $B_1$ need to be evaluated, resulting in a substantial time saving. 

%-----------------------------------------------------------------------
% Continuous matter distribution
\subsection{Continuous matter distribution} 

The continuous matter distribution map that results from line intensity or weak lensing, differs from the discrete case because the map does not have the meaning of counts. It can instead be interpreted as a $\delta$-field, where $\delta$ has a meaning of relative over-density:

\begin{align}
\delta(X,Y,Z) = \frac{D(X,Y,Z) - R(X,Y,Z)}{R(X,Y,Z)} \ , 
\label{eq:densitycont}
\end{align}
where $D$ and $R$ are continuous data and random fields defined on a grid, as is the $\delta$-field, which is identical to $W_0$ map. The convolution of $\delta$-field with kernels  produces $W_i$ maps. 
Since  $\delta$-field is already  normalized by the random matter distribution it is not necessary to do it in the $n$-point correlation function, which is equal to:

\begin{align}
\xi_n(s_1, s_2...s_{(n-1)}) =\sum W_0  W_1(s_1)... W_{(n-1)}(s_{(n-1)}) \ . 
 \label{eq:ksin_cont}
\end{align}
The two point correlation function is then

\begin{align}
\xi_2(s_1) = \sum W_0  W_1  \ . 
\label{eq:ksi2_cont}
\end{align} 

%-----------------------------------------------------------------------
% Cross-correlation between different tracers
\subsection{Cross-correlation between different tracers} 

The algorithm can be used to evaluate cross-correlation between different tracers. In this case we have two (possibly background subtracted) matter maps $\Delta_0$ and $\Delta_1$. Then map of weights $W_0$ is equal to $\Delta_0$,  and $K_1$ is convolved with $\Delta_1$ resulting in $W_1$. The two point correlation function is then evaluated according to Eq.~\ref{eq:ksi2_cont} or Eq.~\ref{eq:ksi2_delta} depending on whether the tracer counts are normalized to the random distribution or not. 

%-----------------------------------------------------------------------
% Performance study
\section{Performance study}

We evaluate the performance of {\tt ConKer} using SDSS DR12 CMASS galaxies \citep{ross2017clustering}, their associated random catalogs, and an  ensemble of MultiDark-Patchy mocks \citep{kitaura2016clustering,rodriguez2016clustering}. We apply {\tt ConKer} to the SGC and NGC catalogs for data, randoms, and 40 mocks. In this performance study, we used the following values for the cosmological parameters: $c = 300000$ km$/$s, $H_0 = 100h$ km$/$s$/$Mpc, $\Omega_{M} = 0.307$, $\Omega_{\Lambda} = 0.693$, and $\Omega_{k} = 0$. This choice was motivated by the cosmological parameters used in the generation of the mock catalogs. For this study, which highlights the algorithm's ability to probe correlations near the BAO scale, we compute correlation functions for a distance range of $54.8\mbox{---}139.8$ $h^{-1}$Mpc in 17 bins of width of $5$ $h^{-1}$Mpc. In all cases, the standard systematic \citep{ross2017clustering} as well as FKP weights \citep{feldman1993power} were used to create the density field map. The grid spacing for each run (with the exception of the timing study discussed in \S~\ref{subsec:timing}) is $g_s = 5$ $h^{-1}$Mpc. 

%-----------------------------------------------------------------------
% Distance calibration
\subsection{Distance calibration}

\begin{figure}
\includegraphics[width=\linewidth]{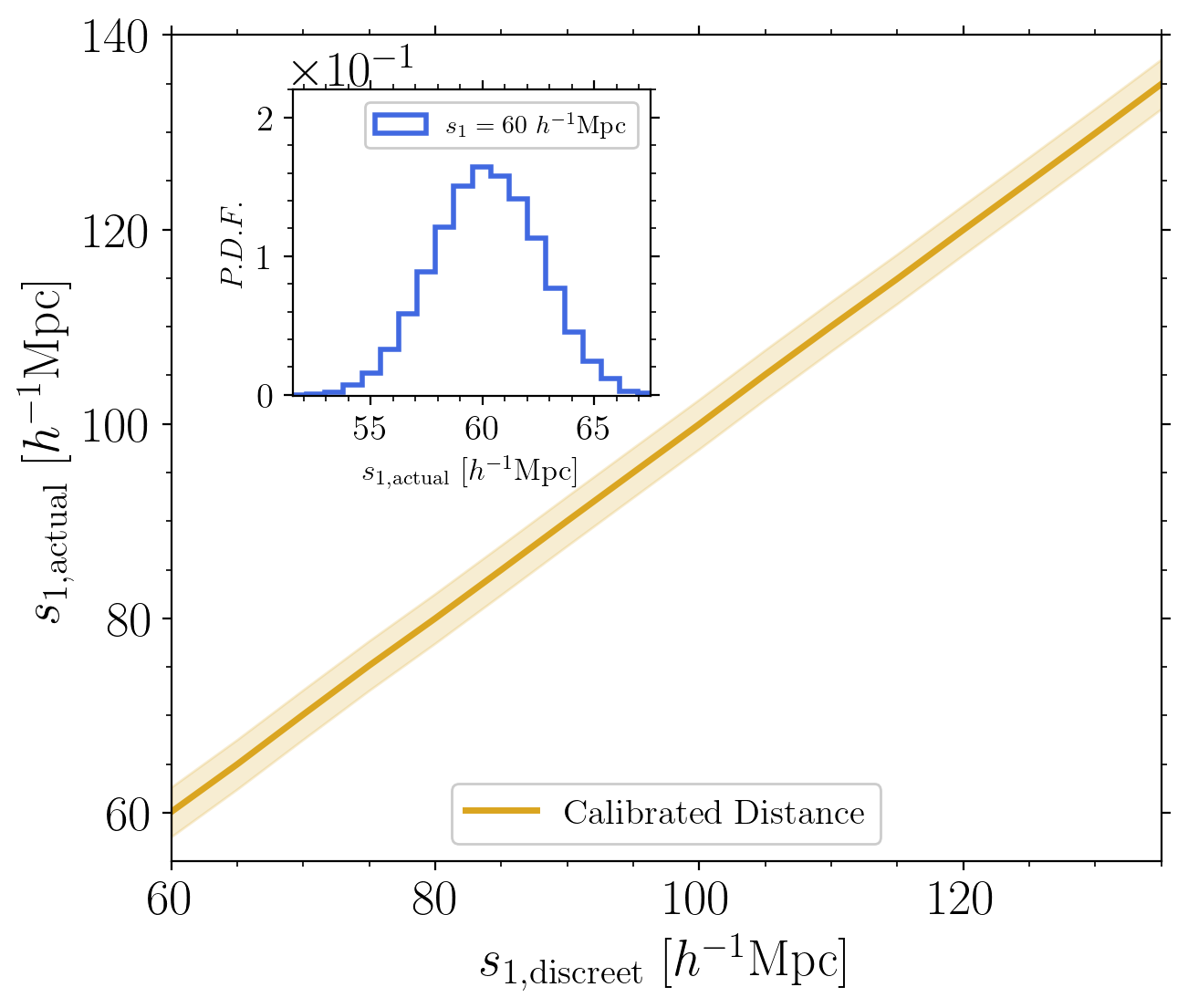}
\caption{The true distances between the tracers and the center of the kernel vs the discretized estimation of the distance in {\tt ConKer} (gold). The shaded region gives the RMS of the true distance distribution. The inset shows one such distribution for a kernel with size $s_1 = 60$ $h^{-1}$Mpc.}
\label{fig:calib}
\end{figure}

As in any algorithm that uses a map of tracers defined on a grid, there is a certain degree of degradation in the precision of the distance between the tracers. It is important to quantify the precision of distance determination. We calibrate the distance using a subsample of a random catalog, where we compare the distance measured on the grid with the one determined by a brute force calculation. Figure~\ref{fig:calib} shows the dependence of the mean true  distance on the kernel size. The shaded region correspond to the RMS of the distribution in the true distance given a particular value of the kernel size, which is shown in the inset for a kernel size of 60 $h^{-1}$Mpc. The RMS is approximately 2.5 $h^{-1}$Mpc, which corresponds to one half grid spacing. The calibration procedure must be performed using a random catalog for each value of the grid spacing, $g_s$. 

%-----------------------------------------------------------------------
% Timing study
\subsection{Timing study}
\label{subsec:timing}

We present the analytical considerations of the time complexity of the three stages of the algorithm: \textit {mapping}, \textit {convolution} and \textit {summation} in Appendix~\ref{app:time}. Here we show the results of the timing study  performed using a personal computer with a 2.3 GHz Intel Core i7 processor and 8 GB of memory. All execution times are in units of CPU seconds. 

\begin{figure}
\includegraphics[width=\linewidth]{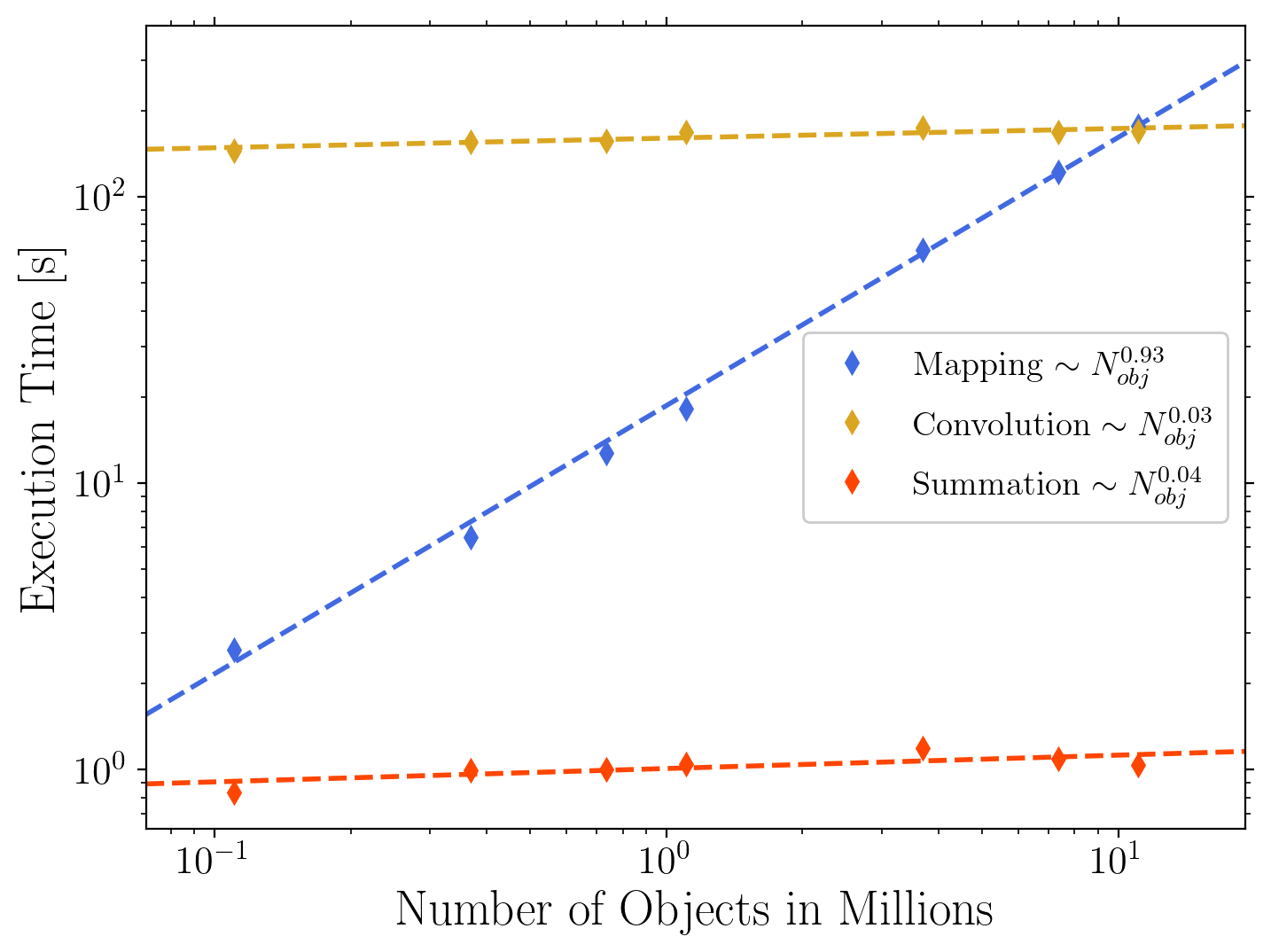}
\includegraphics[width=\linewidth]{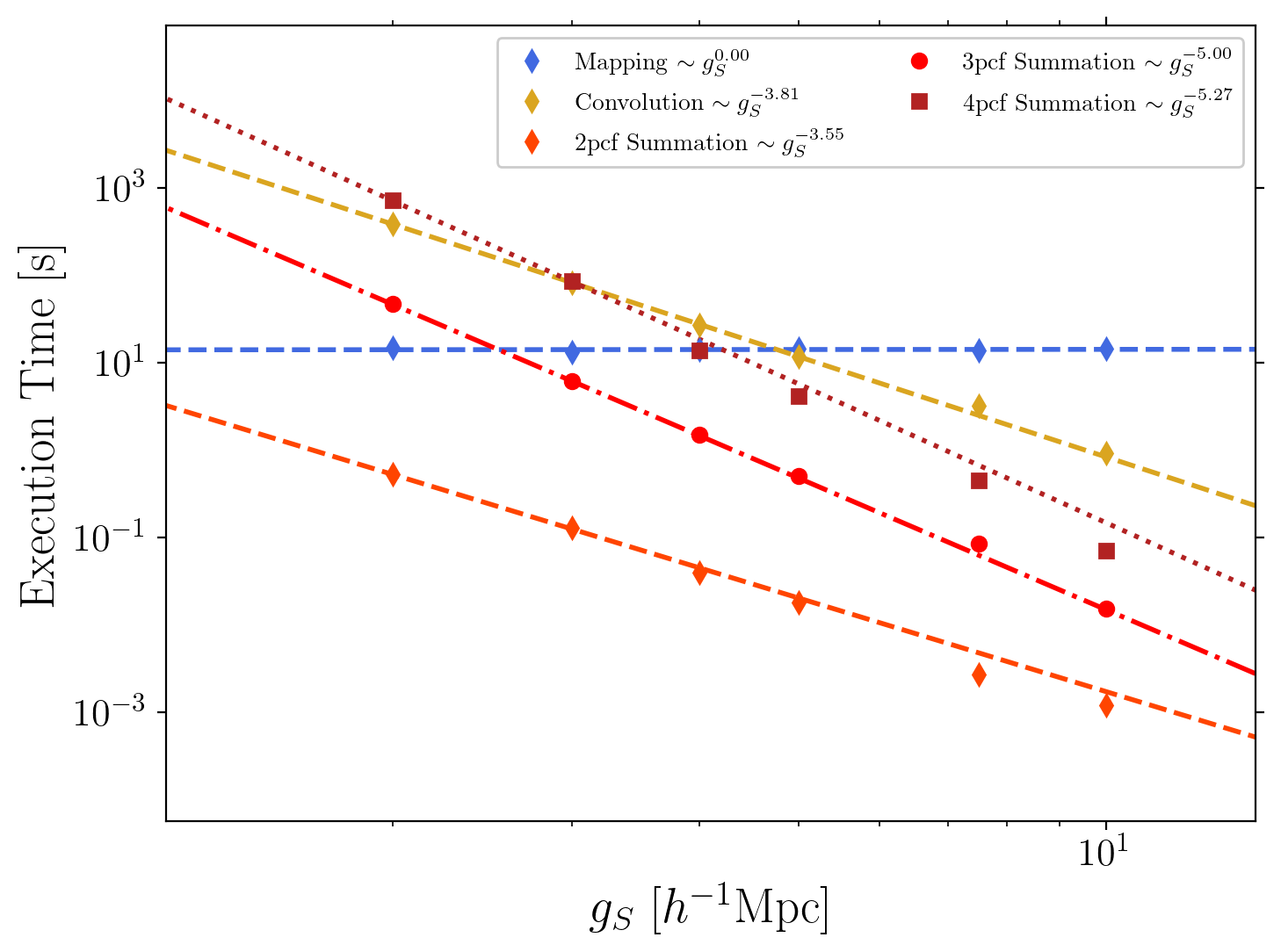}
\caption{Color inline. The execution time of {\tt ConKer} as applied to subsamples of the SDSS DR12 CMASS SGC galaxies. Top: the dependence on the number of combined data and random objects, using a grid spacing $g_S = 5$ $h^{-1}$Mpc. Bottom: the dependence on the grid spacing, $g_S$, for 0.87M objects. Points present the measured time. Lines are the results of the fit to a power law, with the scaling given in the legend.
}
\label{fig:timingSplit}
\end{figure}

The primary advantage of {\tt ConKer} is in the behavior of the execution time as a function of the number of objects, $N$, shown in top plot of Fig.~\ref{fig:timingSplit} for the three stages of the algorithm. The  surveyed volume, $V$ is kept fixed. As expected the execution time of both \textit {convolution} and \textit {summation} are independent of $N$. \textit {Mapping} is a $\mathcal{O}(N)$ calculation, and starts dominating for catalogs larger than 10M objects. 

The main parameter that determines the computation time of {\tt ConKer} is the number of grid cells, $N_c$, which depends the grid spacing, $g_s$ (see the bottom plot of Fig~\ref{fig:timingSplit} for $n=2,3,4$).  \textit {Mapping} is independent of $g_s$. For 2pcf \textit {convolution}  is expected to scale as $g_s^{-4} \log g_s$, and \textit {summation} as $g_s^{-4}$. Both parts scale somewhat slower than expected. The execution time of \textit {convolution} is independent of the correlations order $n$. Each additional order of correlation is expected to add one power of $g_s$ to \textit {summation} time. This is roughly in agreement with the observed dependence for 3pcf and 4pcf. \textit {Summation} starts dominating the execution time for 4pcf for grid spacing below 2 $h^{-1}$Mpc. For  $n>4$ the execution time is expected to be dominated by \textit {summation}, which is a $O(N_c^{(n+2)/3})$ calculation. For lower $n$ or higher grid spacing \textit {convolution} is the dominant part,  expected to be $O(g_s^{-4}\log g_s )$, or $O(N_c^{4/3} \log N_c)$. 
 
The algorithm's memory requirements are discussed in Appendix~\ref{app:memory}.

%-----------------------------------------------------------------------
% Comparison to nbodykit
\subsection{Comparison to nbodykit}

We present a comparison of the 2pcf and 3pcf evaluated using {\tt ConKer} and {\tt nbodykit} \citep{nbodykit}. The latter uses more traditional pair-counting methods, and may be used to evaluate the performance of {\tt ConKer}. In {\tt nbodykit}, the 2pcf is defined according to the estimator of Landy and Szalay \citep{Landy:1993yu,Hamilton:1993fp}:

\begin{align}
\hat{\xi_2}(s) = \frac{DD(s) - 2DR(s) + RR(s)}{RR(s)} \ ,
\label{eq:2pcfdef}
\end{align}
where $DD$, $RR$, and $DR$ are the normalized distributions of the pairwise combinations of galaxies from data, $D,$ and random, $R,$ catalogs (plus cross terms) at given distances $s$ from one another. The {\tt nbodykit} 3pcf is defined according to \cite{Slepian:2015qza} and we consider the $\ell = 0$ (isotropic) case. This corresponds to the {\tt ConKer} 3pcf.

Using {\tt nbodykit}, the 2pcf and 3pcf are calculated for the SGC sample of the SDSS DR12 CMASS galaxies and randoms. Due to the restriction that the 3pcf algorithm in {\tt nbodykit} scales as $\mathcal{O}(N^2)$ \citep{Slepian:2015qza}, where $N$ is the number of objects, the random catalog was reduced to 1M objects. Full random catalogs were used when running {\tt ConKer}. The data catalog remained unchanged in both cases. The {\tt ConKer} 3pcf was completed in $350s$ and the {\tt nbodykit} 3pcf in $9870s$.

\begin{figure}
\includegraphics[width=\linewidth]{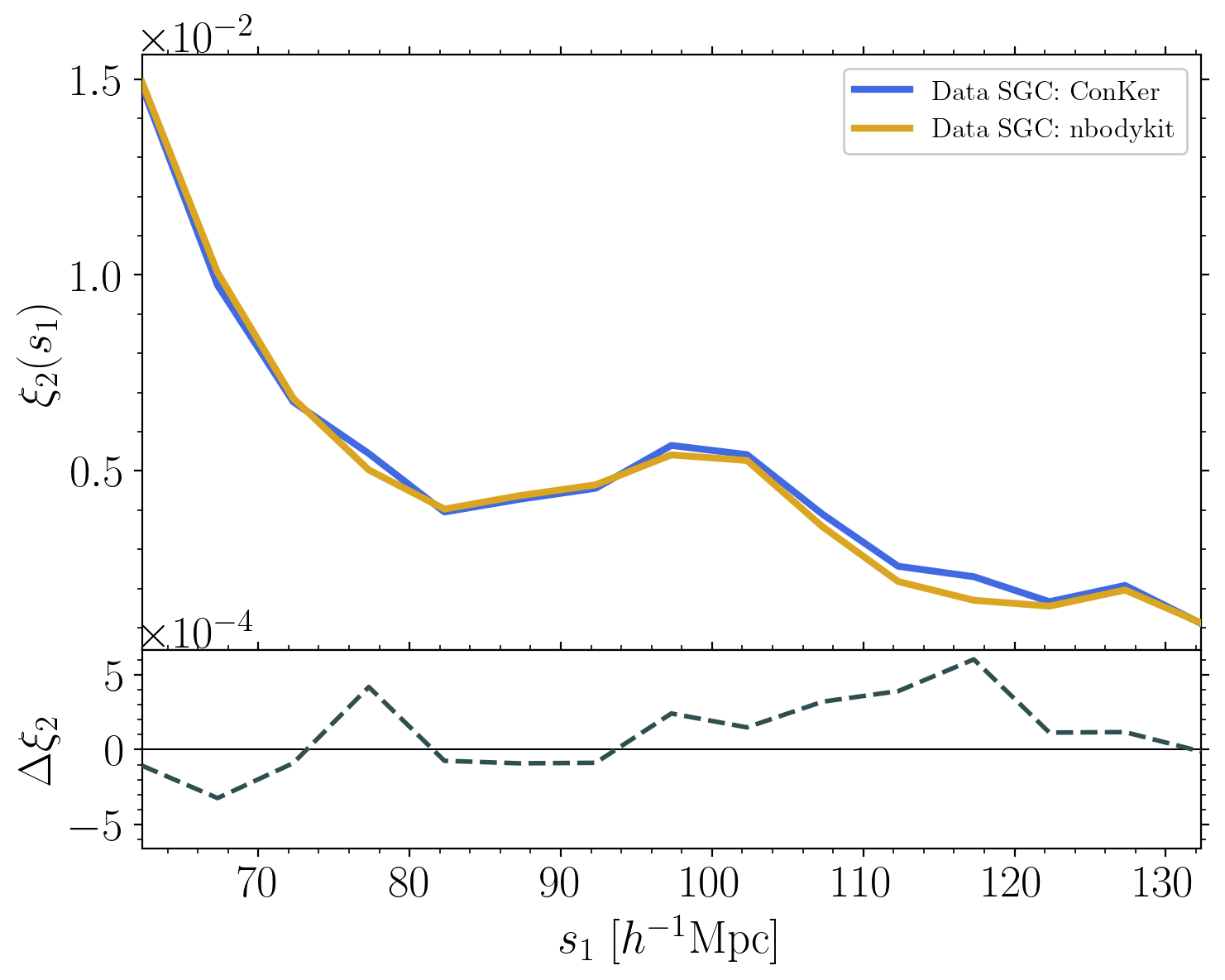}
\caption{The top panel shows the 2pcf, $\xi_2(s_1)$, for the CMASS galaxy sample of the SDSS DR12 survey computed using {\tt ConKer} (blue) and the 2pcf computed using {\tt nbodykit} (gold). The bottom panel gives the residual between the two, defined as $\xi_{2,{\tt ConKer}} - \xi_{2,{\tt nbodykit}}$ (grey dashed line). The solid black line in the bottom panel shows $\Delta\xi_2 = 0$.}
\label{fig:comparison2pt}
\end{figure}

The comparison between the two methods of computing the 2pcf is given in Fig.~\ref{fig:comparison2pt}. The two algorithms agree with one another across the entire range of $s_1$, and the residual between the 2pcf using {\tt ConKer} and {\tt nbodykit} is around an order of magnitude lower than the the value of $\xi_2$. The observed small differences are due to the precision of the distance determination on the grid in {\tt ConKer} resulting in some bin-to-bin migrations, while {\tt nbodykit} calculates the actual distance between the two points. 

\begin{figure}
\includegraphics[width=\linewidth]{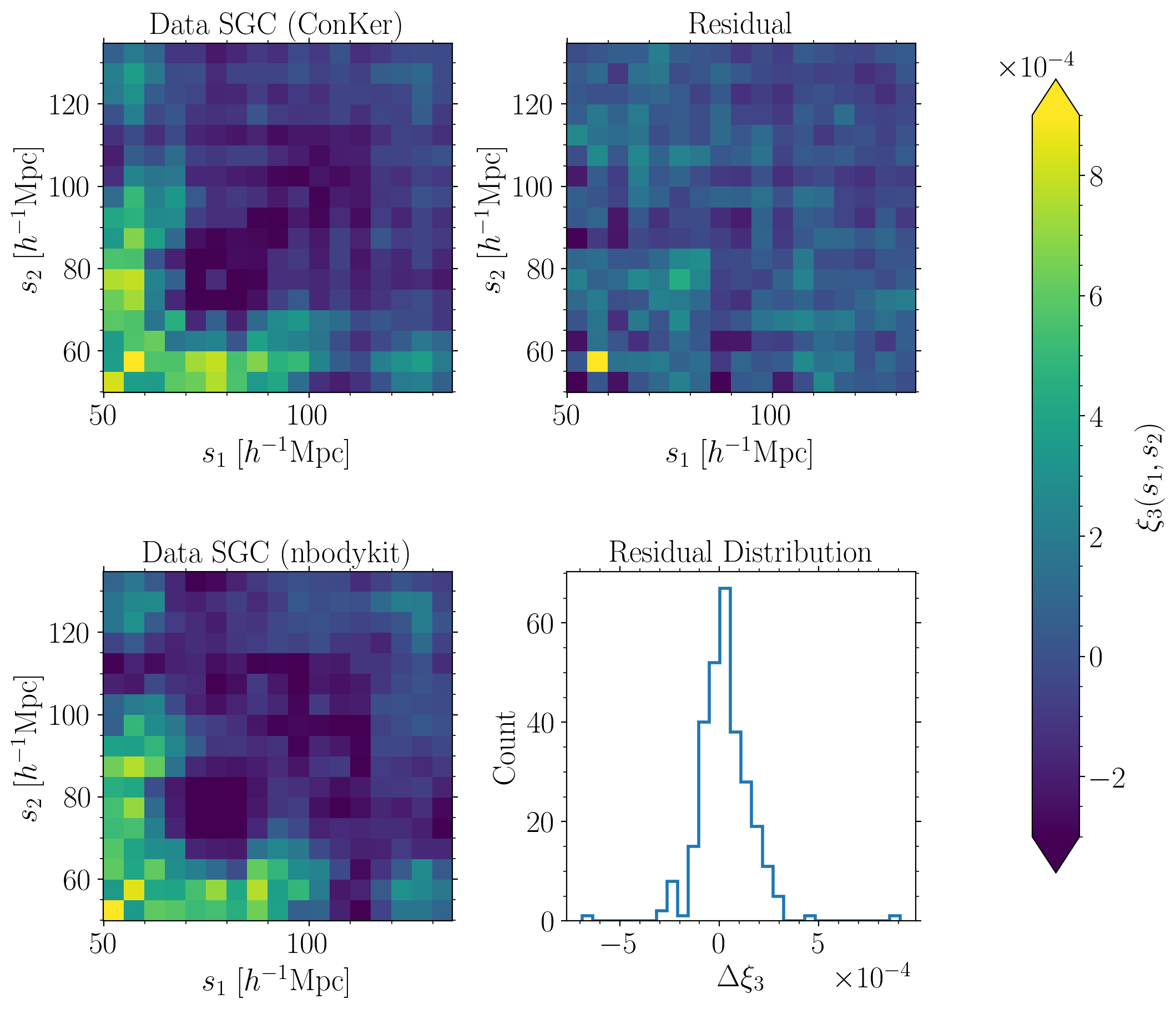}
\caption{The top left panel shows the 3pcf, $\xi_3(s_1,s_2)$, for the CMASS galaxy sample of the SDSS DR12 survey computed using {\tt ConKer} and the bottom left panel shows the 3pcf computed using {\tt nbodykit}. The top right panel gives the residual between the two, defined similarly as $\xi_{3,{\tt ConKer}} - \xi_{3,{\tt nbodykit}}$. The color scale corresponds to all three color panels. The bottom right panel gives a distribution over the residuals. The RMS of this distribution is  $1.34 \times 10^{-4}$.}
\label{fig:comparison3pt}
\end{figure}

We repeat this procedure with the on and off diagonal elements of the 3pcf (Fig.~\ref{fig:comparison3pt}). Here, both methods agree as well, as demonstrated by the residual between the two methods shown in the bottom right of Fig.~\ref{fig:comparison3pt}. Some difference is due to the  reduction in the size of the randoms catalog for the {\tt nbodykit} calculation, which leads to larger statistical fluctuations. Overall, {\tt ConKer} is in a good agreement with standard algorithms, and has the unique advantage of being able to probe higher order correlations as well, with relatively little increase in run-time.

%-----------------------------------------------------------------------
% ConKer npcf
\subsection{ConKer npcf}
\label{subsec:conkernpcf}

We use {\tt ConKer} to compute the diagonal elements of the $n$pcf for $n=2,3,4,5$ and off diagonal elements of the 3pcf for the SGC and NGC catalogs. The combination of NGC and SGC catalogs is done by adding together terms in the numerator and denominator of Eq.~\ref{eq:ksin_delta}. For example, the combined 3pcf is given by

\begin{align}
\xi_3(s_1,s_2) =\frac{\sum (W_0 W_1 W_2)_{\mathrm{SGC}} + (W_0 W_1 W_2)_{\mathrm{NGC}}} {\sum (B_0 B_1 B_2)_{\mathrm{SGC}} + (B_0 B_1 B_2)_{\mathrm{NGC}}}\ . 
\label{eq:ksi3combined}
\end{align}

\begin{figure}
\includegraphics[width=\linewidth]{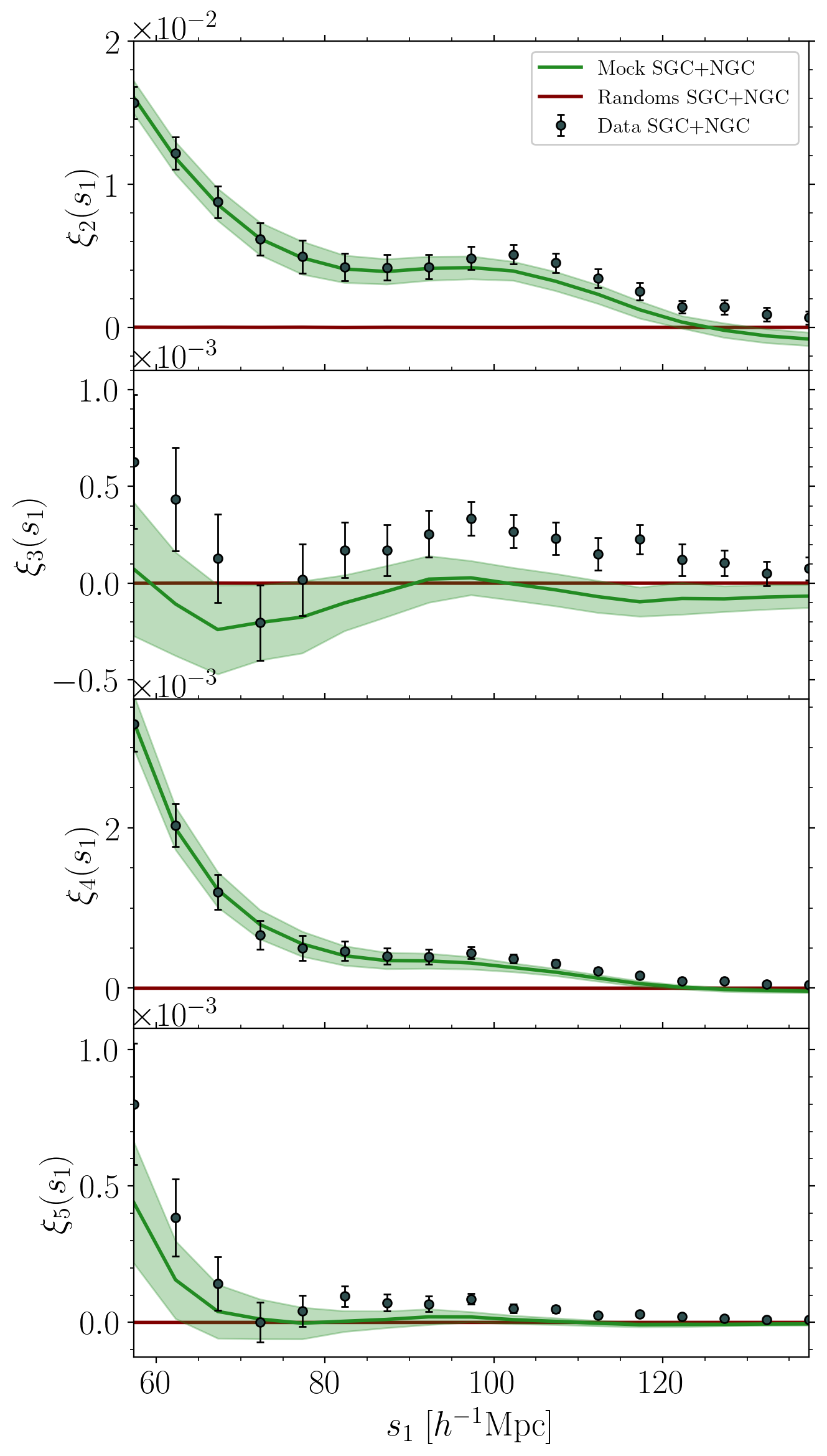}
\caption{The diagonal $n$pcf, $\xi_n(s_1)$, from $n=2$ to $n=5$ (top to bottom panels), for the combined SGC+NGC CMASS galaxy sample of the SDSS DR12 survey (black), the average of the ensemble of MD Patchy mocks (green) and the associated randoms (red). The green shaded region gives the standard deviation across mock catalogs, which is also used to generate the error bars associated with the data catalogs.}
\label{fig:npcfPanel}
\end{figure}

The diagonal elements of the $n$pcf's $n=2, 3, 4$ and $5$ are shown in Fig.~\ref{fig:npcfPanel}. We observe the expected features of the $n$pcf's based on both mock and data catalogs. These include an increase in magnitude at small scales, corresponding to galaxy clustering, and a well defined `bump' at the BAO scale. The $n$pcf's based on the random catalog fluctuate about $\xi_n = 0$ at several orders of magnitude below the signal observed in mocks and data.

\begin{figure}
\includegraphics[width=\linewidth]{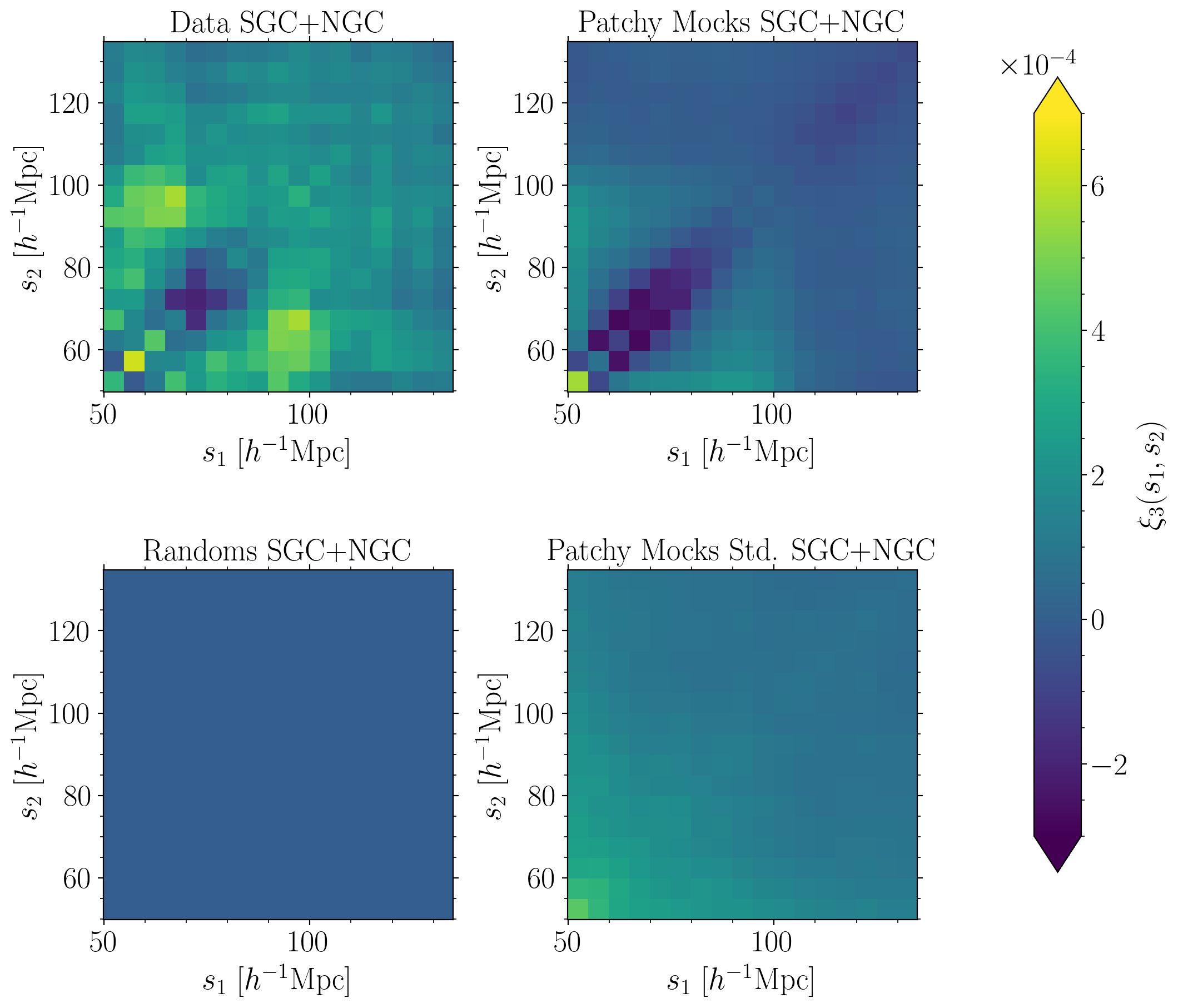}
\caption{The 3pcf calculated based on data (top left), random (bottom left), and mock (top right) catalogs, using the combined SGC and NGC samples. The panel in the bottom right shows the standard deviation across the mock ensemble. The color scale corresponds to all four panels.}
\label{fig:ConKer3ptSQ}
\end{figure}

Both on and off diagonal elements of the 3pcf are shown in Fig.~\ref{fig:ConKer3ptSQ}. The diagonal elements in Fig.~\ref{fig:ConKer3ptSQ} directly correspond to the values in the second panel of Fig.~\ref{fig:npcfPanel}. Off the main diagonal, we observe an increase in the magnitude of $\xi_3$ around $s_1 = 100$ and $s_2 = 60$ $h^{-1}$Mpc for data and mocks. The randoms show no clustering or BAO signal and fluctuate about $\xi_3 = 0$. 

%-----------------------------------------------------------------------
% Discussion
\section{Discussion}

%-----------------------------------------------------------------------
% Algorithmic framework
\subsection{Algorithmic framework}

The query for a pattern in matter distributions may prompt an employment of machine learning techniques.  {\tt ConKer}, being a spatial statistics algorithm  offers an alternative to such approach that is fast and transparent. It exploits the fact that the full set of equidistant points from any given point makes a sphere, with its surface density being a direct measure of how spherically structured this subspace of  points is. Aggregating and combining this measure over the whole space of $N$ objects allows us to calculate the space's $n$ point correlation function.

The algorithm exploits an intrinsic spatial proximity characteristic in the objective of querying for structures of negligible dimensions in a much bigger space. This spatial proximity factor leads to space partitioning algorithms targeting a nearest neighbor query approach, see for example the tree-based $n$pcf algorithms in \cite{MarchThesis}. However, {\tt ConKer} uses this fact as a heuristic in limiting its query space immediately to only the defined separation for each point in the space. Note how this is in contrast with the former technique family. In a nearest neighbor approach to the $n$pcf problem, the query space is grown at each point in the greater embedding space, aggregating the $n$-point statistic until the greater space is fully queried, whereas {\tt ConKer} aggregates the statistic over all embedding space, and then grows the query space before repeating.

This design choice realized in {\tt ConKer}'s core subroutine, a convolution of the query space with the embedding space performed by an $FFT$ algorithm, distills the complexity from the brute force $\mathcal{O}(N_c^2)$ to  
$\mathcal{O}(N_clogN_c)$, where $N_c$ is the total number of grid cells in a discrete 3-dimensional map. This approach, combined with the heuristic above, lets the dominant components of {\tt ConKer} achieve independence of the number of objects as shown in Fig. \ref{fig:timingSplit}. There is certainly a trade-off between the sparsity of the whole space and the bias towards linear complexity in number of objects, as expected from an $FFT$-based algorithm, but even for very dense catalogs, we expect the scaling in the number of objects to be capped by $\mathcal{O}(N)$, where $N$ is the total number of objects. Ultimately, {\tt ConKer} is a hybrid algorithm that draws from both computational geometry and signal processing to achieve linear complexity in the number of objects.

%-----------------------------------------------------------------------
% Applications beyond correlation functions
\subsection{Applications beyond correlation functions}

\begin{figure}
\includegraphics[width=\linewidth]{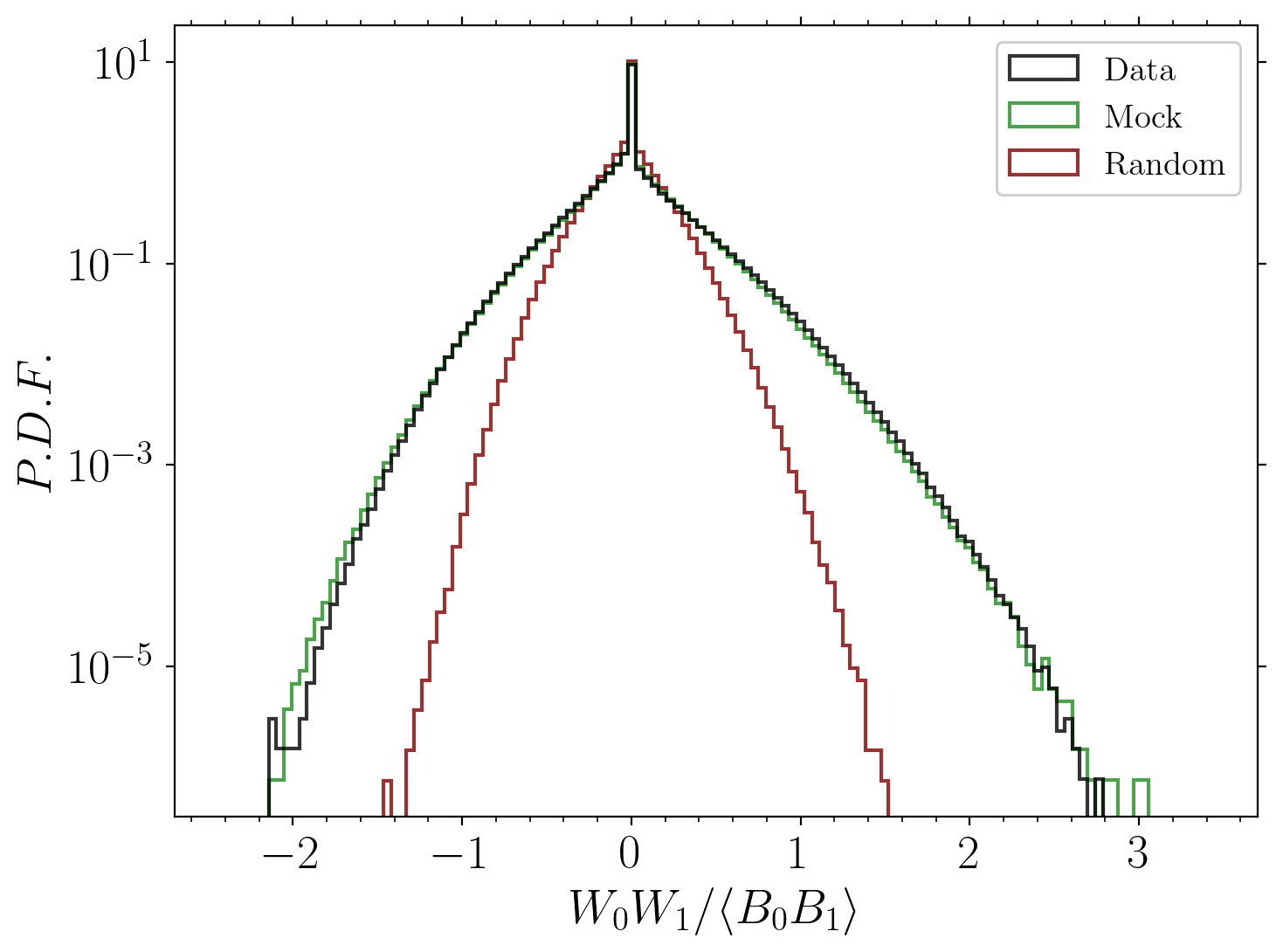}
\caption{The distribution, $W_0 W_1 / \langle B_0 B_1 \rangle$ at $s_1 = 107.3$ $h^{-1}$Mpc, for the CMASS galaxy sample of the SDSS DR12 survey (black), one of the MD Patchy mocks (green) and the associated randoms (red). This plot includes SGC galaxies only.}
\label{fig:wishbone}
\end{figure}

Though traditionally the order of correlation $n$ is viewed as an important parameter, for the equidistant $n$pcf, all the information is entirely encoded by the weights $W_0$ and $W_1$. It was pointed out in \cite{carron2012} that the $n$pcf is inadequate in capturing the tails of non-Gaussianities. The distributions over $W_0$ and $W_1$ (as opposed to their sum over the sample as is used in the $n$pcf) could recover that sensitivity, which is a subject for future studies. The distribution of the product of two weights,  $W_0W_1$ normalized by the average $B_0B_1$ is presented in Fig.~\ref{fig:wishbone} for the kernel size of 107.3 $h^{-1}$Mpc for data, mock and random catalogs. 

\begin{figure}
\includegraphics[width=\linewidth]{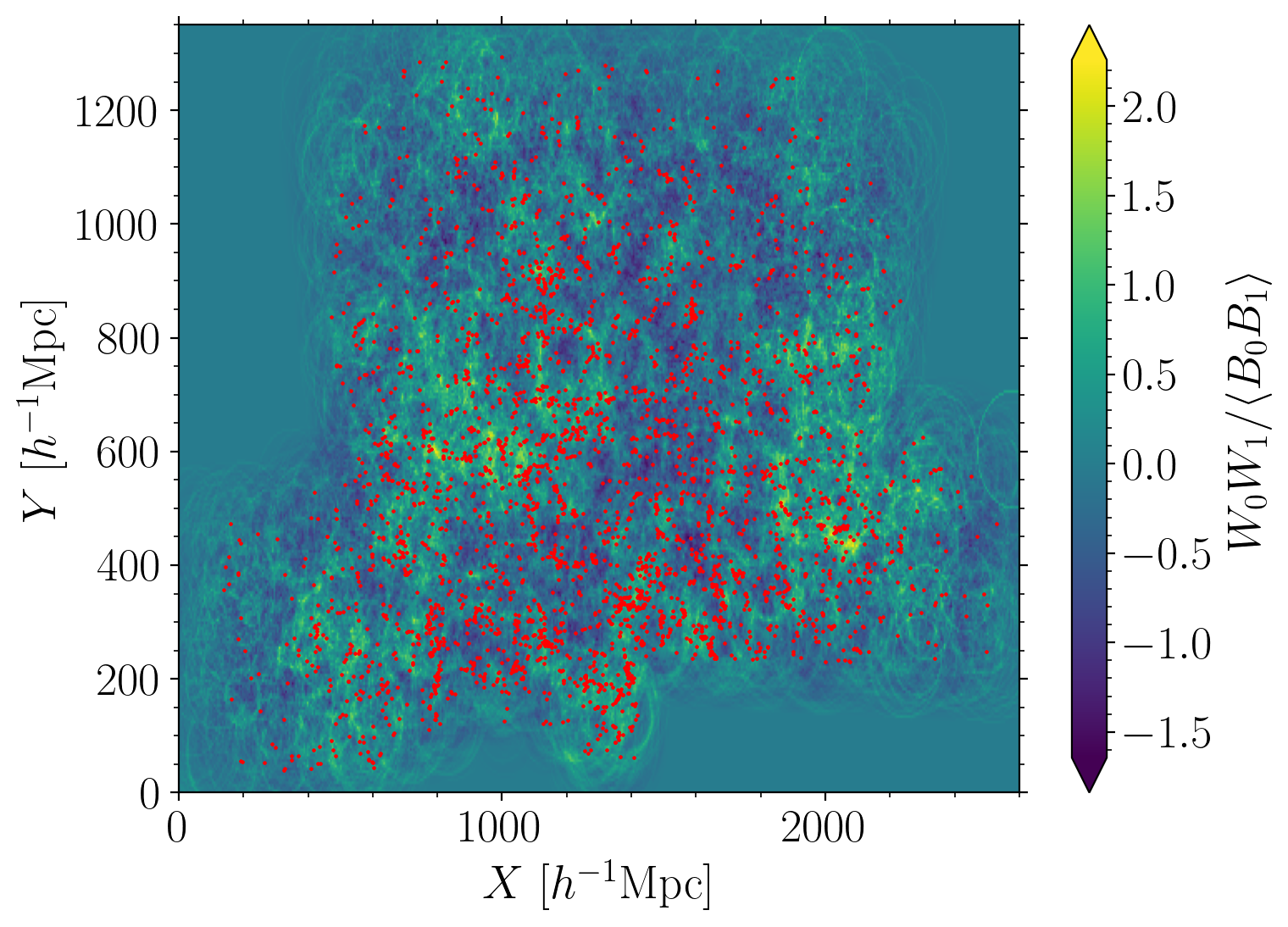}
\caption{A 5 $h^{-1}$Mpc thick slice of the SDSS DR12 SGC volume, where the color scale shows the value of $W_0 W_1 / \langle B_0 B_1 \rangle$ at $s_1 = 107.3$ $h^{-1}$Mpc, and the red stars indicate the positions of galaxies.}
\label{fig:skyslice}
\end{figure}

Moreover, $W_0$ and $W_1$ as well as their product are maps. While in the $n$pcf, the location information is entirely lost, in the maps produced by {\tt ConKer}, it is preserved and can be used for cross correlation studies between different tracers, such as weak lensing, Ly$\alpha$ and CMB. The preservation of this spatial information is illustrated by Fig.~\ref{fig:skyslice}, where the DR12 SGC galaxies are plotted on top of the values of $W_0 W_1 / \langle B_0 B_1 \rangle$ for a kernel size approximately equal to the BAO scale. The figure corresponds to a slice of thickness 5 $h^{-1}$Mpc. The observed circles around the boundaries of the catalog or the sparsely populated regions are a feature of the algorithm, since a single galaxy contributes to $W_1$ everywhere at a distance $s$ from the galaxy  forming a sphere. The highest values  in the map (yellow color scale) correspond to the regions that are both densely populated and have several galaxies displaced by a given scale from this point, as one would expect from the centers of the baryon acoustic oscillations. Similarly, lowest values (blue color scale) correspond to under-dense regions and troughs of BAO originating from this location. Such an aggregated approach gives access to the matter distribution in the early universe. 

%-----------------------------------------------------------------------
% Conclusions
\section{Conclusion}

We presented {\tt ConKer}, an algorithm that convolves spherical kernels with matter maps allowing for fast evaluation of the isotropic $n$-point correlation functions. The algorithm can be broken into three stages: \textit { mapping, convolution} and \textit {summation}. Execution time of \textit {convolution} and \textit {summation} are independent of the catalog size, $N$, while \textit {mapping} is a $\mathcal{O}(N)$ calculation, which starts dominating for catalogs larger than 10M objects. For the correlation orders $n<5$,  \textit {convolution} is the dominant part with complexity $\mathcal{O}(N_c^{4/3} \log N_c)$, where $N_c$ is the number of the grid cells. For higher $n$ the overall time complexity is dominated by \textit {summation}, which is $\mathcal{O}(N_c^{(n+2)/3})$. 

Comparison to the standard techniques shows a good agreement. We study the performance using SDSS DR12 CMASS galaxies, their associated random catalogs, and an  ensemble of MultiDark-Patchy mocks. The results up to $n=5$ are presented. Further metrics that may offer additional sensitivity to primordial non-Gaussianities are also suggested (e.g. distribution over weights $W_i$).

%-----------------------------------------------------------------------
% Acknowledgements
\begin{acknowledgements}

The authors would like to thank Z. Slepian for his interest and insightful comments, F. Weisenhorn for assistance with the {\tt nbodykit} 3pcf calculations, and S. BenZvi, K. Douglass and S. Gontcho A Gontcho for useful discussions. The authors acknowledge support from the U.S. Department of Energy under the grant DE-SC0008475.0. 

Funding for SDSS-III has been provided by the Alfred P. Sloan Foundation, the Participating Institutions, the National Science Foundation, and the U.S. Department of Energy Office of Science. The SDSS-III web site is \url{http://www.sdss3.org/}. SDSS-III is managed by the Astrophysical Research Consortium for the Participating Institutions of the SDSS-III Collaboration including the University of Arizona, the Brazilian Participation Group, Brookhaven National Laboratory, Carnegie Mellon University, University of Florida, the French Participation Group, the German Participation Group, Harvard University, the Instituto de Astrofisica de Canarias, the Michigan State/Notre Dame/JINA Participation Group, Johns Hopkins University, Lawrence Berkeley National Laboratory, Max Planck Institute for Astrophysics, Max Planck Institute for Extraterrestrial Physics, New Mexico State University, New York University, Ohio State University, Pennsylvania State University, University of Portsmouth, Princeton University, the Spanish Participation Group, University of Tokyo, University of Utah, Vanderbilt University, University of Virginia, University of Washington, and Yale University.

The massive production of all MultiDark-Patchy mocks for the BOSS Final Data Release has been performed at the BSC Marenostrum supercomputer, the Hydra cluster at the Instituto de Fısica Teorica UAM/CSIC, and NERSC at the Lawrence Berkeley National Laboratory. We acknowledge support from the Spanish MICINNs Consolider-Ingenio 2010 Programme under grant MultiDark CSD2009-00064, MINECO Centro de Excelencia Severo Ochoa Programme under grant SEV- 2012-0249, and grant AYA2014-60641-C2-1-P. The MultiDark-Patchy mocks was an effort led from the IFT UAM-CSIC by F. Prada’s group (C.-H. Chuang, S. Rodriguez-Torres and C. Scoccola) in collaboration with C. Zhao (Tsinghua U.), F.-S. Kitaura (AIP), A. Klypin (NMSU), G. Yepes (UAM), and the BOSS galaxy clustering working group.

\end{acknowledgements}

%-----------------------------------------------------------------------
% References
\bibliographystyle{aa}
\bibliography{sample}

\begin{thebibliography}{20}
\expandafter\ifx\csname natexlab\endcsname\relax\def\natexlab#1{#1}\fi

\bibitem[{Acquaviva {et~al.}(2003)Acquaviva, Bartolo, Matarrese, \&
  Riotto}]{Acquaviva:2002ud}
Acquaviva, V., Bartolo, N., Matarrese, S., \& Riotto, A. 2003, Nucl. Phys. B,
  667, 119

\bibitem[{Bartolo {et~al.}(2004)Bartolo, Komatsu, Matarrese, \&
  Riotto}]{Bartolo:2004if}
Bartolo, N., Komatsu, E., Matarrese, S., \& Riotto, A. 2004, Phys. Rept., 402,
  103

\bibitem[{Brown {et~al.}(2021)Brown, Mishtaku, Demina, Liu, \&
  Popik}]{brown2021algorithm}
Brown, Z., Mishtaku, G., Demina, R., Liu, Y., \& Popik, C. 2021, Astronomy \&
  Astrophysics, 647, A196

\bibitem[{Carron \& Neyrinck(2012)}]{carron2012}
Carron, J. \& Neyrinck, M. 2012, ApJ, 750, 28 (2012), 28, 750

\bibitem[{Demina {et~al.}(2018)Demina, Cheong, BenZvi, \&
  Hindrichs}]{demina2018computationally}
Demina, R., Cheong, S., BenZvi, S., \& Hindrichs, O. 2018, Monthly Notices of
  the Royal Astronomical Society, 480, 49

\bibitem[{Feldman {et~al.}(1993)Feldman, Kaiser, \& Peacock}]{feldman1993power}
Feldman, H.~A., Kaiser, N., \& Peacock, J.~A. 1993, arXiv preprint
  astro-ph/9304022

\bibitem[{Hamilton(1993)}]{Hamilton:1993fp}
Hamilton, A. J.~S. 1993, Astrophys. J., 417, 19

\bibitem[{Hand(2018)}]{nbodykit}
Hand, N. e.~a. 2018, Astron.J., 4, 156

\bibitem[{Kitaura {et~al.}(2016)Kitaura, Rodr{\'\i}guez-Torres, Chuang, Zhao,
  Prada, Gil-Mar{\'\i}n, Guo, Yepes, Klypin, Sc{\'o}ccola,
  {et~al.}}]{kitaura2016clustering}
Kitaura, F.-S., Rodr{\'\i}guez-Torres, S., Chuang, C.-H., {et~al.} 2016,
  Monthly Notices of the Royal Astronomical Society, 456, 4156

\bibitem[{Landy \& Szalay(1993)}]{Landy:1993yu}
Landy, S.~D. \& Szalay, A.~S. 1993, Astrophys. J., 412, 64

\bibitem[{Maldacena(2003)}]{Maldacena:2002vr}
Maldacena, J.~M. 2003, JHEP, 05, 013

\bibitem[{March(2013)}]{MarchThesis}
March, W.~B. 2013, PhD thesis, Georgia Institute of Technology

\bibitem[{Meerburg {et~al.}(2019)}]{Meerburg:2019qqi}
Meerburg, P.~D. {et~al.} 2019 [\eprint[arXiv]{1903.04409}]

\bibitem[{Philcox {et~al.}(2021)Philcox, Slepian, Hou, Warner, Cahn, \&
  Eisenstein}]{philcox2021encore}
Philcox, O. H.~E., Slepian, Z., Hou, J., {et~al.} 2021, ENCORE: Estimating
  Galaxy $N$-point Correlation Functions in $\mathcal{O}(N_{\rm g}^2)$ Time

\bibitem[{Rodr{\'\i}guez-Torres {et~al.}(2016)Rodr{\'\i}guez-Torres, Chuang,
  Prada, Guo, Klypin, Behroozi, Hahn, Comparat, Yepes, Montero-Dorta,
  {et~al.}}]{rodriguez2016clustering}
Rodr{\'\i}guez-Torres, S.~A., Chuang, C.-H., Prada, F., {et~al.} 2016, Monthly
  Notices of the Royal Astronomical Society, 460, 1173

\bibitem[{Ross {et~al.}(2017)Ross, Beutler, Chuang, Pellejero-Ibanez, Seo,
  Vargas-Magana, Cuesta, Percival, Burden, S{\'a}nchez,
  {et~al.}}]{ross2017clustering}
Ross, A.~J., Beutler, F., Chuang, C.-H., {et~al.} 2017, Monthly Notices of the
  Royal Astronomical Society, 464, 1168

\bibitem[{Slepian \& Eisenstein(2015)}]{Slepian:2015qza}
Slepian, Z. \& Eisenstein, D.~J. 2015, Mon. Not. Roy. Astron. Soc., 454, 4142

\bibitem[{Slepian \& Eisenstein(2016)}]{Slepian:2015qwa}
Slepian, Z. \& Eisenstein, D.~J. 2016, Mon. Not. Roy. Astron. Soc., 455, L31

\bibitem[{Yuan {et~al.}(2018)Yuan, Eisenstein, \& Garrison}]{Yuan:2018qek}
Yuan, S., Eisenstein, D.~J., \& Garrison, L.~H. 2018, Mon. Not. Roy. Astron.
  Soc., 478, 2019

\bibitem[{Zhang \& Yu(2011)}]{zhang2011FFT}
Zhang, X. \& Yu, C. 2011, Third IEEE International Conference on Cloud
  Computing Technology and Science, pp, 634

\end{thebibliography}

%-----------------------------------------------------------------------
% Appendix 
\appendix
\section{Algorithm requirements}

%-----------------------------------------------------------------------
\subsection{Estimation of time complexity}
\label{app:time}

Let us define the relevant parameters:
\begin{itemize}
\item $N$ - number of objects in catalog (since random catalogs are larger, their size  is the leading contribution);
\item $n$ - order of the correlation;
\item $g_s$ - grid spacing;
\item $V$ - volume of a Cartesian box that fully encloses 
the surveyed region;
\item $N_c=V/g_s^3$ - number of grid cells;
\item $s$ - separation of galaxies probed;
\item $s_{max}$ - maximum separation of galaxies probed;
\item $N_s = s_{max}/g_s = s_{max} (N_c/V)^{1/3}$ - number of steps in $s$;
\item $V_k=s^3$ - kernel volume for galaxy separation $s$;
\item $N_k=V_k/g_s^3$ - number of grid cells in kernel volume.
\end{itemize}
Here we present the evaluation of the time complexity of the three stages of the algorithm: \textit { mapping, convolution} and \textit {summation}.
\begin{enumerate}
\item \textit { Mapping - $O(N)$}

Since $N$ objects must be placed in grid cells, the time complexity of the \textit { mapping} is proportional to $N$:
\begin{align}
t_1 \propto N\, .
\label{eq:t1}
\end{align}

It does not depend on neither $n$, $N_c$, nor $N_k$. 

\item \textit { Convolution - $O(N_c^{4/3} \log N_c)$} 

\textit { Convolution} of cubic volumes of the size $N_k$ is performed  using $FFT$ \footnote{Typical complexity of an $FFT convolution$ is $N\log N$.} $N_c$ times at each step in $s$. Hence the complexity of each step is $O(N_c\log N_k)$.  The kernel size itself depends on $s$. We can put an upper bound on it at $s_{max}$. The time complexity of the \textit{convolution} is:

\begin{align}
t_2 \propto N_s N_c \log N_{k} < \frac{s_{max}}{g_s} \frac{V}{g_s^3} \log \frac{V_k}{g_s^3} \propto g_s^{-4} \log g_s \propto N_c^{4/3} \log N_c\, .
\label{eq:t2}
\end{align}
We note that the time complexity of \textit{ convolution} does not depend neither on $n$, nor on $N$, and is the same for on and off diagonal elements. 
\item \textit { Summation: full -$O(N_c^{(n+2)/3} )$,  diagonal - $O(N_c^{4/3})$}

At each one of the $N_s$ steps a product of $n$ grids, each of dimensionality $n$ with each side having a length of $\sqrt[3]{N_c}$ is evaluated, assuming cubic grids. Most of these products are identical.  The number of unique combinations calculated is $\begin{pmatrix}N_s \\ n-1 \end {pmatrix}$, which contains $(n-1)$ terms. Hence the time complexity of \textit { summation} is

\begin{align}
t_3 \propto \begin{pmatrix}N_s \\ n-1 \end {pmatrix} n N_c \propto n \big(\frac{s_{max}}{g_s}\big)^{(n-1)} \frac{V}{g_s^3}  \propto g_s^{-(n+2)} \propto N_c^{(n+2)/3} \, .
\label{eq:t3}
\end{align}

For a special case of evaluating only the diagonal elements of $n$pcf the algorithm takes a product of $n$ arrays each $N_c$ long $N_s$ times, hence the time complexity of \textit { summation} is

\begin{align}
t_3^{diag} \propto  n N_c N_s \propto n \frac{s_{max}}{g_s} \frac{V}{g_s^3}  \propto g_s^{-4} \propto N_c^{4/3} \, .
\label{eq:t3diag}
\end{align}
We note that for a general case for $n>2$\textit { summation} has the highest complexity and becomes dominant for very  low values of grid spacing. For the diagonal elements the complexity of this stage is $N_c^{4/3}$, or lower than that of the \textit { convolution} stage.

\end{enumerate}

%-----------------------------------------------------------------------
\subsection{Estimation of memory requirements}
\label{app:memory}

A run of {\tt ConKer} necessitates several discrete maps that span the survey volume to be held on RAM. For instance, calculating the 2pcf of SDSS DR12 CMASS SGC galaxies with a grid spacing of $5$ $h^{-1}$Mpc produces maps of approximate dimensions ($400^3$). Assuming a 64-bit underlying architecture and grids populated by single-precision floating point numbers, such a contiguous block of memory is just under a Gigabyte. A full run on SGC galaxies needs an available run time memory in the lower single digits in Gigabytes. This memory requirement increases for larger catalogs, but only in proportion to the additional survey volume. 

{\tt ConKer} also requires more memory when computing the off-diagonal elements of higher order $n$-point correlations as opposed to computing just diagonal elements. This is due to the fact that for off-diagonal correlations, the algorithm needs to store the grids corresponding to the convolution of the density grid with each kernel size in the scan, $K_i$, until the very end of the scan, whereas for a diagonal correlation it simply discards the grids at each consecutive scale. Since this algorithm is purposed for use on calculating the $n$pcf at discrete separation values with the total number of such steps being in the low double digits, memory scaling for the off-diagonal case remains feasible on any modern personal computer.

%-----------------------------------------------------------------------
\end{document}